\documentstyle[amscd,12pt,psamsfonts]{amsart}

\newtheorem{defn}{Definition}
\newtheorem{thm}[defn]{Theorem}

\newtheorem{cor}[defn]{Corollary}
\newtheorem{lem}[defn]{Lemma}
\newtheorem{prop}[defn]{Proposition}

\theoremstyle{remark}
\newtheorem{rem}{Remark}
\theoremstyle{remark}
\newtheorem{exam}{Example}

\numberwithin{equation}{section}
\numberwithin{defn}{section}

\begin{document}


\newcommand\aut{\operatorname{Aut}}
\renewcommand\char{\operatorname{char}}
\renewcommand\det{\operatorname{Det}}      
\newcommand\detd{\operatorname{Det}^\ast}
\newcommand\der{\operatorname{Der}}
\newcommand\ext{\operatorname{Ext}}
\newcommand\grv{{\operatorname{Gr}}(V)}
\newcommand\grb{\operatorname{Gr}^\bullet}
\newcommand\gr{\operatorname{Gr}}
\newcommand\glv{{\operatorname{Gl}}(V)}
\newcommand\glve{{\widetilde{\operatorname{Gl}}(V)}}
\newcommand\gl{\operatorname{Gl}}
\newcommand\lieglv{{\operatorname{gl}}(V)}
\renewcommand\hom{\operatorname{Hom}}
\newcommand\id{\operatorname{Id}}
\newcommand\im{\operatorname{Im}}
\newcommand\Lie{\operatorname{{\frak L}ie}}
\newcommand\limi{\varinjlim}
\newcommand\limil[1]{\underset{#1}\varinjlim\,}
\newcommand\limp{\varprojlim}
\newcommand\limpl[1]{\underset{#1}\varprojlim\,}
\newcommand\ord{\operatorname{ord}}
\newcommand\proj{\operatorname{Proj}}
\newcommand\spk{{\operatorname{Spec}}(k)}
\renewcommand\sp{\operatorname{Spec}}       
\renewcommand\sf{\operatorname{Spf}}       
\newcommand\res{\operatorname{Res}}
\newcommand\tr{\operatorname{Tr}}
\newcommand\vir{\operatorname{Vir}}

\renewcommand\o{{\mathcal O}}      
\renewcommand\L{{\mathcal L}}       
\renewcommand\c{{\mathcal C}}  
\renewcommand\P{{\mathbb P}}    
\newcommand\Z{{\mathbb Z}}    
\newcommand\A{{\mathbb A}}    
\newcommand\M{{\mathcal M}}
\newcommand\m{{\frak m}}
\newcommand\B{{\mathcal B}}
\newcommand\R{{\mathbb R}}    
\newcommand\C{{\mathbb C}}

\newcommand\dobleto{{\small\begin{gathered}
\raisebox{-4pt}{$\longrightarrow$} \\
\raisebox{5pt}{$\longrightarrow$}\end{gathered}}}

\newcommand\iso{@>{\sim}>>}
\newcommand\kz{\underline{k((t))}^*}
\newcommand\beq{       
      \setcounter{equation}{\value{defn}}\addtocounter{defn}1
      \begin{equation}}

\renewcommand{\thesubsection}{\thesection.\Alph{subsection}}

\title []  {Automorphism Group of $k((t))$: \\ Applications to the
Bosonic String}

\author[J. M. Mu\~noz \and F. J. Plaza ] 
{J. M. Mu\~noz Porras \\
\and \\  F. J. Plaza Mart\'{\i}n\thanks{} \\ 
\medskip\tiny Departamento de Matem\'aticas \\ Universidad de Salamanca}

\address{ Departamento de Matem\'aticas \\ Universidad de Salamanca
\\ Plaza de la Merced 1-4\\ Salamanca 37008. Spain.}

\thanks{
\noindent {\it E-mail}: jmp@@gugu.usal.es \\ {\it E-mail}:
fplaza@@gugu.usal.es
\\ This work is partially supported by the
CICYT research contract n. PB96-1305 and Castilla y Le\'on regional
goverment contract SA27/98.}


\date\today

\maketitle

\begin{abstract}
This paper is concerned with the formulation of a non-pertubative
theory of the bosonic string. We introduce a formal group $G$ which we
propose as the ``universal moduli space'' for such a formulation. This is
motivated because $G$ establishes a natural link between representations
of the Virasoro algebra and the moduli space of curves.
\end{abstract}



\setcounter{tocdepth}1
\tableofcontents

\section{Introduction}

On the moduli space of smooth algebraic curves of genus
$g$,  $\M_g$, one can define a family of determinant invertible sheaves
$\{\lambda_n\vert n\in\Z\}$. In a remarkable paper, Mumford (\cite{Mu})
proved the existence of canonical isomorphisms:
$$\lambda_n\,\iso\,\lambda_1^{(6n^2-6n+1)}\qquad \forall n\in\Z$$
which have been studied in depth from different approaches. 

For instance,
within the frame of string theory, these isomorphisms are one of the main
tools in the explicit computation of the Polyakov measure for bosonic
strings in genus $g$ (\cite{BK,MM}). Proposals for developing a
genus-independent (or ``non-pertubative'') formulation of the theory of
bosonic strings have been made by several authors (e.g.
\cite{BR,Mo,BNS}).

In this paper we propose a  ``universal moduli space'' as the
main ingrident for a non-perturbative string theory which is different
from those introduced by the above authors.

Following the spirit of previous papers (\cite{AMP,MP}), where 
a ``formal geometry'' of curves and Jacobians was developed (see
\cite{BF,Pl} for other applications of these ideas), we introduce a formal
group scheme
$G$ representing the functor of automorphisms of $k((t))$ (see
\S~\ref{sect:G}); more precisely, the points of $G$ with values in a
$k$-scheme $S$ are:
$$G(S)\,=\, \aut_{H^0(S,\o_S)-\text{alg}}H^0(S,\o_S)((t))$$

The formal group scheme $G$ might be interpreted as a formal moduli scheme
for parametrized formal curves. The canonical action of $G$ on the
infinite Grassmannian $\gr(k((t))dt^{\otimes n})$ allow us to construct an
invertible sheaf, $\Lambda_n$, on $G$ (for every $n\in\Z$) endowed with
a bitorsor structure. Using a generalization of the Lie Theory for certain
non commutative groups (given in Appendix \ref{sect:Lie}), we prove that
these sheaves satisfy an analogous formula of the Mumford Theorem; that is,
there exist canonical isomorphisms (see Theorem \ref{thm:locmum}):
$$\Lambda_n\,\iso\,\Lambda_1^{(6n^2-6n+1)}\qquad \forall n\in\Z$$

To show that our formula  is a local version of Mumford's, rather than a
mere ``coincidence'', we relate
$G$ and the moduli of curves by means of infinite Grassmannians (see
subsection~\ref{subsec:modcur} for precise statements). Let
$\M_g^\infty$ be the moduli space of pointed curves of genus $g$
with a given parameter at the point (see Definition
\ref{5:defn:modinf}). Then, the action of $G$ on
$\gr(k((t)))$ induces an action, $\phi$, on $\M_g^\infty$. Moreover, given
a rational point $X\in \M_g^\infty$, the action induces a morphism of
schemes:
$$G\,\overset{\phi_X}\longrightarrow \M_g^\infty$$
Let $\widehat \phi_X$ be the composite of the immersion of $\widehat G$
(the formal completion of $G$ at the identity) into $G$, $\phi_X$, and the
projection $\M_g^\infty\to\M_g$. Let $(\M_G^\infty)_X^{\widehat{\,}}$ be the
formal completion of $\M_G^\infty$ at $X$. Then, from the surjectivity of
the map $\widehat G\to (\M_G^\infty)_X^{\widehat{\,}}$ induced by
$\phi_X$ (see Theorem \ref{5:thmmain}), it follows easily that there exist
isomorphisms:
$$\widehat\phi_X^*(\lambda_n)\,\iso\, \Lambda_n\qquad \forall n\in\Z$$

Finally, the last section offers a proposal on how to apply these results
to a non-perturbative formulation of the bosonic string. The explicit
development of these ideas and the geometric interpretation of partition
functions in terms of the geometry of the group $G$ will be performed
elsewhere.

\section{Background on Grassmannians}
\subsection{The Grassmannian $\gr(k((t)))$}

This section summarizes results on infinite Grassmannians as given
in \cite{AMP} in order to set notations and to recall the facts we
will need.

Below, $V$ will always denote the $k$-vector space
$k((t))$ and $V^+$ the subspace $k[[t]]$. Let $\B^f$ be the set of
subspaces generated by $\{t^{s_0},t^{s_1},\ldots\}$ for every
strictly increasing sequence of integers $s_0<s_1<\ldots$ such that
$s_{i+1}=s_i+1$ for $i>>0$. Let $\B$ denote the set of subspaces of
$V$ given by the $t$-adic completion of the elements of $\B^f$. We can now
interpret $\B$ as a basis of a topology on $V$. It is easy to characterize
the neighborhoods of $0$ as the set of subspaces $A$ of $V$ such that
there exists an integer $n>>0$ with
$t^nk[[t]]\subseteq A$ and it is of finite codimension. 

Now the pair $(V,\B)$ satisfies the following properties:
\begin{itemize}
\item the topology is separated and $V$ is complete,
\item for every $A,B\in\B$, it holds that $(A+B)/(A\cap B)$ is
finite dimensional,
\item if $A,B\in\B$, then $A+B,A\cap B\in\B$,
\item $V/A=\limil{B\in\B}( B+A/A)$ for every $A\in\B$.
\end{itemize}
and hence there exists a $k$-scheme, called the
Grassmannian of $(V,\B)$ and denoted by $\grb(V)$, whose $S$-valued
points is the set: {\small $$\left\{
\begin{gathered}
\text{quasi-coherent sub-$\o_S$-modules $L\subseteq \hat V_S$ such
that for every point $s\in S$,}\\ L_{k(s)}\subseteq \hat
V_{k(s)}\text{ and there exists an open neighborhood $U$ of $s$ and
}A\in\B\\
\text{such that }\hat V_U/L_U+ \hat A_U=(0) \text{ and }L_U\cap
\hat A_U\text{ is free of finite type}
\end{gathered}
\right\}$$} ($k(s)$ is the residual field of $s$) where $\hat
L_T:=\limp (L/L\cap A_S)\otimes_{\o_S}\o_T$ for a submodule $L$ of
$V_S$ and a morphism of $k$-schemes $T\to S$.

The very construction of $\grb(V)$ shows that
$\{F_A\;\vert\;A\in\B\}$ is an open covering by affine subschemes
where $F_A$ is the
$k$-scheme whose $S$-valued points are:
$$\left\{
\text{locally free sub-$\o_S$-modules $L\subseteq \hat V_S$ such
that }L_S\oplus \hat A_S\simeq\hat V_S
\right\}$$

From this fact one deduces (see \cite{AMP}) that the complexes of
$\o_{\grb(V)}$-modules $\L\oplus \hat A_{\grb(V)}\to \hat V_{\grb(V)}$ are
perfect ($\L$ being the universal object of $\grb(V)$) for
every $A\in\B$. Moreover, the Euler-Poincar\`e
characteristic of the complex $\L\oplus \hat A_{\grb(V)}\to \hat
V_{\grb(V)}$:
$$L \longmapsto \dim(L\cap V^+)-\dim(V/L+V^+)$$
 gives the decomposition of
$\grb(V)$ into connected components. The connected component of
characteristic
$0$ will be denoted by $\grv$. It is easy to show that these
complexes are all quasi-isomorphic.

From the theory of \cite{KM} on determinants, it
follows that their determinants are well defined and that they are
isomorphic. The choice of $V^+\in\B$ now enables us to construct a
line bundle on the Grassmannian as follows: on the connected
component of characteristic $n$ consider the determinant
of $\det( \L\oplus t^n\hat V^+_{\gr^n(V)}\to
\hat V_{\gr^n(V)})$. The resulting bundle will be called ``the
determinant bundle'' and will be denoted simply by $\det_V$. 
 
It is also known that given a complex $\L\oplus \hat
A_{\grb(V)}\overset{\delta_A}\to \hat V_{\grb(V)}$ ($A\in\B$), the
morphism $\delta_A$ gives a section of $\det(\L\oplus \hat
A_{\grb(V)}\to \hat V_{\grb(V)})^*$. By fixing the basis $\{t^n\vert
n\in\Z\}$ of $V$ one checks  that the induced isomorphisms among
determinants of these complexes are compatible (see
\cite{AMP}). Using such isomorphisms the above-defined section gives
a section $\Omega_A$ of $\detd_V$. The section defined on the connected component of characteristic $n$ by the determinant of the addition homomorphism  $\L\oplus t^n\hat V^+_{\gr^n(V)}\to\hat V_{\gr^n(V)}$ will be denoted by $\Omega_+$.

\subsection{The Linear Group $\glv$}

For each $k$-scheme $S$, let us denote by
$\aut_{\o_S}(\hat V_S)$ the group of automorphisms of the
$\o_S$-module $\hat V_S$.

\begin{defn}\hfill
\begin{itemize}
\item A sub-$\o_S$-module $L\subseteq \hat V_S$ is said to be a
$\B$-neighborhood if there exists a  vector subspace $A\in\B$ such
that $\hat A_S\subset L$ and $L/\hat A_S$ is locally free of finite
type.
\item An automorphism $g\in\aut_{\o_S}(\hat V_S)$ is called
$\B$-bicontinuous if $g(\hat A_S)$ and $g^{-1}(\hat A_S)$ are
$\B$-neighborhoods for all $A\in\B$.
\item The linear group, $\glv$, of $(V,\B )$ is the contravariant
functor over the category of $k$-schemes defined by:
$$S\rightsquigarrow \glv(S):=\{g\in\aut_{\o_S} (\hat V_S)\text{ such
that $g$ is $\B $-bicontinuous }\}$$
\end{itemize}
\end{defn}

\begin{thm}\label{thm:glact}  
There exists a natural action, $\mu$, of $\glv$ on the
Grassmannian, preserving the determinant bundle.
\end{thm}

\begin{pf} 
The first part is easy to show. It suffices to prove that $g(L)$
belongs to $\grb(V)(S)$ for an $S$-valued point $L\in\grb(V)(S)$ and an
arbitrary $g\in\glv(S)$ using that $g$ is $\B$-bicontinuous.

Note that given $g\in\glv(S)$ and an $S$-scheme, $T$, one
has an induced isomorphism $\hat V_S/\hat A_S\to \hat V_S/g(\hat
A_S)$ for each $A\in\B$. Twisting by $\o_T$, and taking inverse
limit over $A\in \B$, one obtains  an $\o_T$-automorphism $g_T$ of
$\hat V_T$, which due to the very construction is $\B$-bicontinuous.
Moreover, the map:
$$\begin{aligned}
\glv(S)& \to \glv(T) \\ g& \mapsto g_T\end{aligned}$$
 is functorial. So, for an element $g\in\glv(S)$ we have
constructed $g_T\in\glv(T)$ for every $S$-scheme $T$; hence, $g$
yields an $S$-automorphism of $\grb(V)_S:=\grb(V)\times_k S$.
We have then constructed a functor homomorphism:
$$\begin{aligned}
\glv &\to \underline\aut(\grb(V))\\
g&\mapsto g_{\bullet}
\end{aligned}$$
where $\underline\aut(\grb(V))(S):=\aut_{\text{$S$-sch}}(\grb(V)_S)$.

With the expression ``preserving the determinant bundle'' we mean
that $g_{\bullet}^*p_1^*\det\simeq p_1^*\det \otimes p_2^*N$ (where
$p_i$ denotes the projection onto the $i$-th factor of $\grb(V)\times_k
S$) for a line bundle $N$ over $S$. It is therefore enough to prove
the statement when $S$ is a local affine scheme. 

Recall that:
$$g_{\bullet}^*p_1^*\det_V\,\simeq\,
\det\big(g_{\bullet}^*p_1^*\L\oplus g_{\bullet}^*p_1^* \hat A_{\grb(V)}
\to  g_{\bullet}^*p_1^* \hat V_{\grb(V)}\big)$$
for $A\in\B$. Take $A\in\B$ such  that  $\hat A_S\subseteq
g^{-1}(\hat V^+_S)$ and
$g^{-1}(\hat V^+_S)/\hat A_S$ are free of finite type. Then, $g$
induces an isomorphism:
$$g_{\bullet}^*p_1^*\det_V\simeq p_1^*\det_V \otimes
\det\big(p_1^*\hat V^+_{\grb(V)}/g_{\bullet}^*(p_1^*\hat
A_{\grb(V)})\big)^*$$

From the very
construction of $g_{\bullet}$ it follows that there is an
isomorphism:
$$p_1^*\hat
V^+_{\grb(V)}/g_{\bullet}^*(p_1^*\hat A_{\grb(V)})\,\simeq\,
p_2^*\big(\hat V^+_S/ g(\hat A_S)\big)$$
and the claim follows.
\end{pf}

\begin{thm}\label{thm:glve} 
There exists a canonical central
extension of functors of groups over the category of $k$-schemes:
$$0\to{\Bbb G}_m\to \glve \to \glv \to 0$$ and a natural action,
$\tilde\mu$, of $\glve$ over the vector bundle,
${\Bbb V}(\det_V)$, defined by the determinant bundle lifting the
action $\mu$. 
\end{thm}

\begin{pf}  
For an affine $k$-scheme $S$, define ${\mathcal G}(S)$ as the set of
commutative diagrams (in the category of $S$-schemes):
$$\CD  {\Bbb V}(\detd_V)_S @>{\bar g}>>{\Bbb V}(\detd_V)_S
\\@VVV @VVV \\   \grb(V)_S @>g>>\grb(V)_S
\endCD$$  
where $\bar g$ is an isomorphism and  $g\in\glv(S)$ and the
homomorphism  ${\mathcal G}\to\glv$  by $\bar g\mapsto g$. For an
arbitrary scheme $S$ define ${\mathcal G}(S)$ by sheafication; that is,
consider a covering $\{U_i\}$ by open affine subschemes of $S$ and
${\mathcal G}(S)$ the kernel of the restriction homomorphisms:
$$\prod_i {\mathcal G}(U_i)\,\dobleto\, \prod_{i,j}{\mathcal G}
(U_i\cap U_j)$$

We have then obtained an extension:
$$0\to \prod_{\Z}{\Bbb G}_m\to {\mathcal G}  \to \glv \to 0$$
since $H^0(\grb(V)_S,\o_{\grb(V)_S})=\prod_{\Z} H^0(S,\o_S)$
(\cite{AMP}).

Finally, define $\glve(S)$ as the direct image of this extension by
the morphism $\prod_{\Z}{\Bbb G}_m\to {\Bbb G}_m$ which maps
$\{a_i\}$ to $a_0$. Observe that for any projection 
$\{a_i\}\mapsto a_n$ the resulting extensions are isomorphic.
\end{pf}


Let us compute the cocycle associated with this central extension. For
the sake of clarity we shall begin with the finite dimensional
situation: $V$ finite dimensional, $\{v_1,\ldots, v_d\}$ a basis,
$\B$ consists of all finite dimensional subspaces and
$V^+:=< v_{n+1},\ldots, v_d>$ (for an integer $0\leq n\leq d$). Then,
$\grv$ parametrizes the
$n$-dimensional subspaces of $V$. Let
$\bar g$ denote the morphism
$<v_1,\ldots, v_n>\hookrightarrow V\overset{g}\to V\to V/V^+$ for an
element
$g\in\glv$ (observe that $\bar g$ consists of the first $n$
columns and rows of the matrix associated with $g$).

We now have the following exact sequence:
$$ 0 \to{\Bbb G}_m \to \glve \overset{p}\to \glv \simeq
\aut(\wedge^n V)\to  0 $$

Let us consider the subgroup $\gl^+(V)$
consisting of those automorphisms $g\in\glv$ such that $\bar g$ is
an isomorphism. 
It is easy to check that:
$$ g\,\longmapsto\, \big(g, det(\bar g)\big)$$
is  a section of $p$ over $\gl^+(V)$. The cocycle associated to the central
extension is given by:
$$c(g_1,g_2)\,=\, det \big(\bar g_1\circ (\overline{g_1\circ
g_2})^{-1}\circ \bar g_2\big)$$

The cocyle corresponding to the Lie algebra level follows from a
straightforward computation. Let $\id+\epsilon_i D_i$ be a
$k[\epsilon_i]/\epsilon_i^2$-valued point of $\glv$ ($i=1,2$).
The very definition of the cocycle:
$$c_{\text{Lie}}(D_1,D_2)\epsilon_1\epsilon_2\,=\, c(\id+\epsilon_1
D_1 ,\id+\epsilon_2 D_2)-c(\id+\epsilon_2 D_2,\id+\epsilon_1
D_1)$$
yields the expression:
\beq
c_{\text{Lie}}(D_1,D_2)\,=\tr(D_1^{+-}D_2^{-+}-D_2^{+-}D_1^{-+})
\label{eq:liecocycle}\end{equation} 
 where $D_i^{+-}:V^+\to V^-:=<v_1,\ldots,v_n>$  is
induced by $\id+\epsilon_i D_i\in\glv$ with respect to the
decomposition $V\simeq V^-\oplus V^+$ (and, analogously,
$D_i^{-+}:V^-\to V^+$).

The case of $(V=k((t)),\B,V^+=k[[t]])$ and $V^-=t^{-1}k[t^{-1}]$ is
very similar and the same formulae remain valid.

\section{The Automorphism Group of $k((t))$: $G$}\label{sect:G}

This section aims at studying the functor (on groups) over the
category of $k$-schemes defined by:
$$S\rightsquigarrow
G(S):=\aut_{\text{$H^0(S,\o_S)$-alg}}H^0(S,\o_S)((t))$$ where the
group law in $G$ is given by the composition of automorphisms (here
$R((t))$ stands for $R[[t]][t^{-1}]$ for a commutative ring $R$
with identity; or, what amounts to the same, the Laurent
developments in $t$ with coefficients in $R$). 

\subsection{Elements of $G$}\label{subsect:eleme}

Let us consider the following functor over the category of
$k$-schemes:
$$S\,\rightsquigarrow\, \kz(S):=
\left\{\text{invertibles of } H^0(S,\o_S)((t)) \right\}
$$  

The first result is quite easy to show:

\begin{lem}
The functor homomorphism:
$$\begin{aligned}
\psi_R:\aut_{\text{$R$-alg}}R((t)) &\to \kz(R) \\ g\quad
&\mapsto g(t)
\end{aligned}$$
 induces an injection of $G$ into the connected component of $t$,
$\kz_1$. Moreover, $G(R)\to \kz_1(R)$ is a semigroup
homomorphism with respect to the following composition law on
$\kz_1$:
\beq
\begin{aligned} m:\kz_1(R)\times \kz_1(R) &\to
\kz_1(R) \\ (g(t),h(t))\, &\mapsto h(g(t))
\end{aligned}
\label{3:eq:alaw}\end{equation}
\end{lem}

\begin{thm}\label{3:thm:Gstruct} 
The morphism $\psi_R$ induces a natural isomorphism of functors:
$$G\,\iso\, \kz_1$$
\end{thm}

\begin{pf} 
The only delicate part of the proof is the surjectivity of $\psi_R$. The
idea is to relate $G(R)$ with the group of automorphisms of $R[[x]][y]$.

Let $I$ be the ideal of $R[[x]][y]$ generated by $(x\cdot
y-1)$, and let $\aut_I R[[x]][y]$ be the group:
$$\left\{g\in\aut_{\text{$R$-alg}}R[[x]][y]\text{ such that }
g(I)=I\right\}$$ Since there is an isomorphism
$R[[x]][y]/I\iso R((t))$ (which maps $x$ to $t$ and $y$ to $t^{-1}$),
one has a morphism $\aut_I R[[x]][y]\to
\aut_{\text{$R$-alg}}R((t))$, and a commutative diagram:
$$\CD 
\aut_I R[[x]][y] @>{\bar\psi_R}>> R[[x]][y] @>\pi>> R[[x]][y]/I\\
@VVV @.  @V{\simeq}VV \\
\aut_{\text{$R$-alg}}R((t)) @>\psi_R>> R((t))^*_1 @>>> R((t))^*
\endCD$$ where ${\bar\psi_R}(f):=f(x)$. 

Observe that the induced morphism:
$$\pi\left(
\left\{\begin{gathered}
\text{series $f(x,y)\in x\cdot
R[[x]]\oplus\operatorname{Rad}(R)[y]$}\\
\text{such that the coefficient of $x$ is invertible}
\end{gathered}\right\}\right)
\,\longrightarrow\, R((t))^*_1$$ is surjective. The claim being
equivalent to the surjectivity of
$\psi_R$, it is then enough to show that:
$$\left\{\begin{gathered}
\text{series $f(x,y)\in x\cdot R[[x]]\oplus
\operatorname{Rad}(R)[y]$}\\
\text{such that the coefficient of $x$ is invertible }
\end{gathered}\right\}
\,\subseteq\, \im({\bar\psi_R})$$

Given an element $x\cdot f(x)+n(y)\in x\cdot
R[[x]]\oplus\operatorname{Rad}(R)[y]$ where $f(0)$ is invertible,
consider the following $R$-endomorphism:
$$\begin{aligned}
\phi: R[[x]][y] & \to R[[x]][y] \\ x &\mapsto x\cdot f(x)+n(y) \\ y
&\mapsto \frac{y}{f(x)}\cdot 
\left(1+\frac{y\cdot n(y)}{f(x)}\right)^{-1}
\end{aligned}$$ 
(which is well defined since $f(x)\in R[[x]]^*$ and
$n(y)$ is nilpotent). 

Provided that $\phi$ is an isomorphism, it holds that $\phi(I)=I$ and that 
$\bar\psi_R(\phi)=x\cdot f(x)+n(y)$. To show that $\phi$ is actually an
$R$-isomorphism of $R[[x]][y]$, observe that
$\phi=\phi_3\circ\phi_2\circ\phi_1$ where $\phi_1,\phi_2,\phi_3$ are
$R$-isomorphisms of $R[[x]][y]$ defined by:
{\small
$$
\cases  \phi_2(x)=x\cdot f(x) & \\
 \phi_2(y)=y 
& \endcases
\hfill
\cases \phi_3(x)=x \\
 \phi_3(y)=\frac{y}{f(x)}
\cdot 
\left(1+\frac{y\cdot n(y)}{f(x)}\right)^{-1}
& \endcases
\hfill
\cases \phi_1(x)=x+(\phi_3\circ\phi_2)^{-1}(n(y)) & \\
 \phi_1(y)=y 
& \endcases
$$}

\end{pf}

\subsection{Formal Scheme Structure of $G$}
\label{subsec:k((t))}

Set an $k$-scheme $S$ and an element $f\in\kz(S)$. From
\cite{AMP} we know that the function:
$$\begin{aligned} S &\longrightarrow \Z \\ s &\mapsto v_s(f):=
\text{ order of }f_s\in k(s)((t))
\end{aligned}$$ 
is locally constant and that the connected component of $t$,
$\kz_1$,  is identified with the set of $S$-valued points of a
formal $k$-scheme, $k((t))^*_1$. One therefore obtains an isomorphism
between the functor $G$ and the functor of points of the formal
scheme $k((t))^*_1$:
$$G(S)\,\iso\, k((t))^*_1(S)=\left\{\begin{gathered}
\text{series }(a_{r}t^r+\dots+a_0 +a_1t+\dots)t\text{ such that}\\
a_r,\dots,a_{-1}\in\operatorname{Rad}(R),\, a_0\in R^*\text{
and }r<0 \end{gathered}\right\}$$
(where $R=H^0(S,\o_S)$). 


\subsection{Subgroups of $G$}

Two important subgroups of $G\iso k((t))^*_1$ are the subschemes
$G_+$ and $G_-$ defined by:
$$\begin{gathered}
G_+(S)\,:=\,\left\{t\cdot(1+\sum_{i>0} a_i\,t^i) \text{ where }
a_i\in R\right\}\\
G_-(S)\,:=\,\left\{\begin{gathered}
\text{polynomials }\,t\cdot(a_r\,t^r+\dots+a_1\,t^{-1}+1) \text{
such}
\\
 \text{that }a_i\in R \text{ are nilpotent and $r$ arbitrary}
\end{gathered}\right\}
\end{gathered}
$$
respectively.

Let $\widehat G$ (respectively $\widehat G_-, \widehat {\mathbb G}_m$
and $\widehat G_+$) be the completion of the formal scheme $G$ 
($ G_-, {\mathbb G}_m$ and $ G_+$) at the point
$\{\id\}$.

\begin{lem}
The subgroups $\widehat G_-, \widehat {\mathbb G}_m$ and
$\widehat G_+$ commute with each other and:
$$\widehat G_-\cdot\widehat {\mathbb G}_m\cdot \widehat
G_+\,=\,\widehat G$$
\end{lem}

\begin{pf}
Recall that $\hom(\sp(A),\widehat G)$ is the union of
$\hom(\o/{\frak m}_{\o}^n,A)$ where $\o$ is the ring of $G$ and
${\frak m}_{\o}$ is the maximal ideal corresponding to the
identity. It therefore suffices to show that:
\begin{enumerate}
\item $\widehat G_-(A), \widehat {\mathbb G}_m(A)$ and $\widehat
G_+(A)$ commute with each other,
\item $\widehat G_-(A)\cdot\widehat {\mathbb G}_m(A)\cdot\widehat
G_+(A)\,=\,\widehat G(A)$,
\end{enumerate}
for each local and rational $k$-algebra $A$ such that ${\frak
m}_A^{n+1}=0$ for $n>>0$. 

Let us proceed by induction on $n$. The case $n=1$ is a simple
computation.

\begin{enumerate}
\item Let us prove that $\widehat G_-(A)$ and $ \widehat {\mathbb
G}_m(A)$ commute with each other. Consider the following subgroup
of $G(A)$:
$$H(A)\,:=\,\left\{ a_{n}t^{-n}+\ldots+a_0 \text{ with }a_i\in{\frak
m}_A\text{ for $i<0$ and }a_0\in A^*\right\}$$
and note that we have the group exact sequence:
$$0\to \widehat H(k[{\frak m}^n_A])\to \widehat H(A)\overset{\rho}\to
\widehat H(B)\to 0$$
where $B=A/{\frak m}_A^n$.

For an element $h\in \widehat H(A)$ there exist $h_-\in \widehat
G_-(B)$ and $h_0\in \widehat {\mathbb G}_m(B)$ such that
$\rho(h)=\rho(h_-\circ h_0)$; or what amounts to the same:
$$h_-^{-1}\circ h\circ h_0^{-1}\,\in\, \widehat H(k[{\frak
m}^n_A])$$
The induction hypothesis implies that $\widehat H(k[{\frak m}_A^n])=
\widehat G_-(k[{\frak m}_A^n])\cdot \widehat {\mathbb G}_m(k[{\frak
m}_A^n])$ and hence there exist $h'_-\in \widehat
G_-(k[{\frak m}_A^n])$ and $h'_0\in \widehat {\mathbb G}_m(k[{\frak
m}_A^n])$ such that:
$$h_-^{-1}\circ h\circ h_0^{-1}=
h'_-\circ h'_0$$
and therefore:
$$\widehat H\,=\,\widehat G_-\cdot \widehat {\mathbb G}_m$$
Analogously, one proves that $\widehat H\,=\,
\widehat {\mathbb G}_m\cdot \widehat G_-$.

The proofs of the other commutation relations are similar.

\item Note that $g_0\circ g_-=g_-\circ
g_0$ for $g_0\in \widehat {\mathbb G}_m(A)$ and $g_-\in \widehat
G_-(k[{\frak m}_A^n])$ and proceed similarly.
\end{enumerate}
\end{pf}

\begin{thm}\label{3:thm:Gaction} 
The functor $G$ is canonically a subgroup of $\glv$.
\end{thm}

\begin{pf} 
Note that it suffices to show that $G_-(S),{\mathbb G}_m(S)$ and
$G_+(S)$ are canonically subgroups of $\glv(S)$ for each $k$-scheme
$S$, since:
\begin{itemize}
\item $\widehat G=\widehat G_-\cdot\widehat{\mathbb
G}_m\cdot\widehat G_+$,
\item $\widehat G_-= G_-$ and $\widehat G_+\subseteq G_+$,
\item $G=\widehat G\cdot G_+$.
\end{itemize}

By the very definition of $\glv$, it is enough to prove the case when
$S$ is a local affine scheme, $\sp(R)$. 

The cases of ${\mathbb G}_m$ and $G_+$ are straightforward since:
$$\phi(t^n R[[t]])= t^n R[[t]]\qquad \forall n$$
for $\phi\in {\mathbb G}_m(S)$ or $\phi\in G_+(S)$.

Let us now consider $\phi\in G_-(S)$. Let $u(t)$ be such that 
$\phi^{-1}(t)=t(1+u(t))$. It then holds that:
$$\phi^{-1}(t^r)=t^r(1+u(t))^r
\,=\,t^r\cdot\sum_{i=0}^r\binom{r}{i}u(t)^i$$
Since $u(t)$ is nilpotent, there exists $s$ such that:
$$\phi^{-1}(t^rR[[t]])\,\subseteq\,t^sR[[t]]$$
in other words:
$$t^rR[[t]]\,\subseteq\,\phi(t^sR[[t]])$$

The Nakayama lemma implies that the family $\{\phi(t^s),\ldots,
\phi(t^{r-1})\}$ generates $\phi(t^sR[[t]])/t^rR[[t]]$. Using the fact that
$\phi\in G_-$ one proves that they are linearly independent; summing
up, $\phi(t^sR[[t]])/t^rR[[t]]$ is free of finite type.
\end{pf}

\subsection{The Lie Algebra of $G$, $\Lie(G)$}\label{subsect:lie}

\begin{thm}\label{5:lem4}
There is a natural isomorphism of Lie algebras:
$$\Lie(G)\,\iso \,k((t))\partial_t$$
compatible with their natural actions on the tangent space to
the Grassmannian, $T\grv$. (From now on
$\der_k k((t))$ will denote $k((t))\partial_t$) 
\end{thm}

\begin{pf} 
Take an element $g(t)=t(1+\epsilon g_0(t))\in \Lie(G)$ (recall that
by definition $\Lie(G)=G(k[\epsilon]/\epsilon^2)
\times_{G(k)}\{Id\}$). Let us compute $\mu(g)(t^m)$ for $m\in\Z$:
$$\begin{aligned}
\mu(g)(t^m)&=g(t)^m=t^m(1+\epsilon g_0(t))^m =\\ &=  t^m(1+m\epsilon
g_0(t))=(Id+\epsilon\cdot g_0(t) t\partial_t)(t^m)
\end{aligned}$$  
It is now natural to define the following map:
$$\begin{aligned} \Lie(G) &\to \der_k k((t))\\ t(1+\epsilon
g_0(t))&\mapsto g_0(t) t\cdot\partial_t \end{aligned}$$ 
and this turns out to be an isomorphism of $k$-vector spaces.

In order to check that this map is actually an isomorphism of Lie
algebras, let us compute explicitly the Lie algebra structure of
$\Lie(G)$.

Given two elements $g_n(t)=t(1+\epsilon_1 t^n)$ and
$g_m(t)=t(1+\epsilon_2 t^m)$ (where $\epsilon_i^2=0$), we have: 
$$g_n(g_m(t)) \, =\, g_m(g_n(t)) (1+(m-n)\epsilon_1\epsilon_2
t^{m+n})$$ 
that is:
$$[g_m,g_n]\,=\, (m-n)g_{m+n}$$ 
Since
$[t^{m+1}\partial_t,t^{n+1}\partial_t]=(m-n)\cdot
t^{m+n+1}\partial_t$, one concludes that the map is in fact an
isomorphism of Lie algebras.

Let us check that the actions of these Lie algebras on 
$T\grv$ coincide. Fix a rational point $U\in\grv$ and take an element
$g(t)=t(1+\epsilon g_0(t))\in \Lie(G)$. Clearly, the image of
$(g,U)$ by $\mu$ lies on:
$$T_U\grv \,=\,
\grv(k[\epsilon]/\epsilon^2)\underset{\grv(k)}\times\{U\}
\,\simeq\, \hom_k(U,V/U)$$ 
which is associated with the morphism:
$$U\hookrightarrow V\overset{\cdot t g_0(t)}\longrightarrow
V\to V/U$$
Consider an element $D\in \der_k k((t))$. Then the image of $(D,U)$
under the action of $\der_k k((t))$ on $T\grv$ is:
$$U\hookrightarrow V\overset{D}\longrightarrow
V\to V/U$$
and the conclusion follows.
\end{pf}

Let $\vir$ denote the Virasoro algebra; that is, the Lie algebra
with a basis $\{\{d_m\vert m\in{\mathbb Z}\}, c\}$ and Lie brackets
given by:
$$\begin{aligned}
[d_m,c] &\,=\, 0 \\
[d_m,d_n] & \,=\, (m-n) d_{m+n}+\delta_{n,-m}\frac{(m^3-m)}{12}c
\end{aligned}$$
By abuse of notation $\vir$ and Virasoro will also denote  the Lie
algebra given by $\limpl{n} \vir/\{d_m\vert m>n\}$. Both algebras
have a ``universal'' central extension:
$$\ext^1(k((t))\partial_t,{\mathbb C})\,  =\,{\mathbb C}\cdot \vir$$
and this is the important feature for our approach (see \cite{KR}
Lecture~1, \cite{ACKP}~2.1, \cite{LW}).

\begin{defn}
The central extension of
$G$ given by Theorem \ref{thm:glve}, $\widetilde G$, will be 
called the Virasoro Group.
\end{defn}

\begin{prop}
The Lie algebra of $\widetilde G$, $\Lie(\widetilde G)$, is isomorphic to
the Virasoro algebra, $\vir$.
\end{prop}

\begin{rem}
Let us compute the cocycle associated with $\Lie(\widetilde G)$. Let
$\gl^+(V)$ be the subgroup of $\glv$ consisting of elements $g$ such
that $g(F_{V^+})=F_{V^+}$. Since $\widehat G$ is contained in
$\gl^+(V)$, one can use the formula \ref{eq:liecocycle}. Recall that
a basis of $\Lie(\widehat G)$ is given by the set
$\{g_n(t):=t(1+\epsilon t^n)\vert n\in{\mathbb Z}\}$ since
$\Lie(\widehat G)=\widehat G(k[\epsilon]/\epsilon^2)$.

The element of $\gl^+(V)$ (a ${\mathbb Z}\times {\mathbb Z}$ matrix)
corresponding to $g_m$ is:
$$(g_m)_{ij}=\cases
1 &\text{ if } i=j\\
\epsilon\cdot j &\text{ if } i=j+m\\
0 &\text{ otherwise}
\endcases$$
and the cocycle is therefore:
$$c(g_m,g_n)\,=\,
\delta_{n,-m}\cdot\sum_{j=0}^{n-1}j(j-n)=\delta_{n,-m}\cdot
\frac{m^3-m}{6}$$
\end{rem}

\subsection{Central Extensions of $G$}

We begin with an explicit construction of an important family
of central extensions of $G$.

Fix two integer numbers $\alpha,\beta$ and consider the $k$-vector
space $V_{\alpha,\beta}:=t^\alpha k((t))(dt)^{\otimes \beta}$. The
natural isomorphism:
$$\begin{aligned} d_{\alpha,\beta}:V&\longrightarrow
V_{\alpha,\beta} \\ f(t)&\mapsto t^\alpha f(t)(dt)^{\otimes \beta}
\end{aligned}$$ 
allows us to define a triplet
$(V_{\alpha,\beta},\B_{\alpha,\beta}:=d_{\alpha,\beta}(\B),V^+_{\alpha,\beta}
:=d_{\alpha,\beta}(V^+))$. One has therefore an isomorphism:
$$\grv\,\iso \, \gr(V_{\alpha,\beta})$$

Observe that the action of $G$ on $V_{\alpha,\beta}$ defined by:
{\small $$\big(g(t),t^\alpha f(t) (dt)^{\otimes \beta}\big)\,\mapsto
\, g(t)^\alpha f(g(t))(dg(t))^{\otimes
\beta}=t^\alpha \big(\frac{g(t)}{t}\big)^\alpha
f(g(t))g'(t)^\beta(dt)^{\otimes \beta}
$$ }
induces an action on $\gr(V_{\alpha,\beta})$ (by a
straightforward generalization of Theorem {\ref{3:thm:Gaction}}),
and also in $\grv$:
$$\mu_{\alpha,\beta}:G\times \grv \to \grv$$

Note that $\mu_{0,0}$ is the action of $G$ on $\grv$ defined in the
previous section. Moreover, these actions are related by:
$$\mu_{\alpha,\beta}(g(t))\,=\,
\left(\big(\frac{g(t)}{t}\big)^\alpha\cdot g'(t)^\beta\right)\circ
\mu_{0,0}(g(t))$$ where the first factor is the homothety defined
by itself.

The Theorem {\ref{thm:glve}} implies that there exists a central
extension:
$$0\to {\mathbb G}_m\to \widetilde G_{\alpha,\beta} \to
G\to 0$$
corresponding to the action $\mu_{\alpha,\beta}$. Moreover, it
follows from its proof that $\widetilde G_{\alpha,\beta}$ consists
of commutative diagrams:
$$\CD  {\Bbb V}(\detd_V) @>{\bar g}>>{\Bbb V}(\detd_V) 
\\@VVV @VVV \\   \grv @>\mu_{\alpha,\beta}(g)>>\grv 
\endCD$$  or equivalently:
$$\widetilde G_{\alpha,\beta}\,=\, \{(g,\tilde g) \text{ where }g\in
G\text{ and }
\tilde g:\mu_{\alpha,\beta}(g)^*\det_V\iso\det_V\}$$ since
$\mu_{\alpha,\beta}(g)^*\det_V\simeq\det_V$ for all $g\in G$.
It is not difficult to show that the extensions $\widetilde
G_{\alpha,\beta}$ and $\widetilde G_{\alpha',\beta}$ are isomorphic
for every $\alpha,\alpha'\in\Z$. Then, $\widetilde G_{0,\beta}$
(respectively $\mu_{0,\beta}$) will be denoted by $\widetilde
G_{\beta}$ ($\mu_{\beta}$).
The group law of $\widetilde G_{\beta}$ is:
$$(h,\tilde h)\cdot (g,\tilde g)\,=\, (h\cdot g, \tilde g \circ
\mu_{\beta}(g)^*(\tilde h))$$ since we have:
$$\mu_{\beta}(h\cdot
g)^*\det_V=(\mu_{\beta}(g)^*\circ\mu_{\beta}(h)^*)\det_V
\overset{\mu_{\beta}(g)^*(\tilde h)}\longrightarrow
 \mu_{\beta}(g)^*\det_V\overset{\tilde g}\longrightarrow
\det_V$$

These central extensions induce extensions of the
Lie algebra $\Lie(G)$ whose corresponding cocycles are:
\beq\begin{aligned}
c_{\beta}(m,n)\,&=\,\delta_{n,-m}\cdot
\sum_{j=0}^{n-1}(j+(m+1)\beta)(j-n+(n+1)\beta)
\\
&=\,\delta_{n,-m}\cdot\big( \frac{m^3-m}{6}
\big) (1-6\beta+6\beta^2)
\end{aligned}
\label{eq:lie-mumford}\end{equation}
To obtain such a formula, one only has to check that the matrix
corresponding to $\mu_{\beta}(g_m)$ is:
$$(\mu_{\beta}(g_m))_{ij}=\cases
1 &\text{ if } i=j\\
\epsilon\cdot (j+(m+1)\beta) &\text{ if } i=j+m\\
0 &\text{ otherwise}
\endcases$$

\begin{rem}
It is worth pointing out that one can continue with this geometric
point of view for studying the representations of $\Lie(G)$ since it
acts on the space of global sections of the Determinant line bundle
which contains the ``standard'' Fock space. (For an explicit
construction of sections of $\detd_V$, see \cite{AMP}). An algebraic
study of the representations of $\vir$ induced by
$\mu_{\alpha,\beta}$ has been done in \cite{KR}.
\end{rem}

\subsection{Line Bundles on $G$}\label{subsec:torsors}

Formula \ref{eq:lie-mumford} may be stated in terms of line
bundles. For this goal, let us first recall from \cite{SGA} the
relationships among line bundles, bitorsors and extensions.

Recall that a central extension of the group $G$ by
${\mathbb G}_m$:
$$0\to {\mathbb G}_m \to {\mathcal E} \to G \to 0$$ (${\mathcal E}$
being a group)  determines a bitorsor over $\big(({{\mathbb
G}_m})_G,({{\mathbb G}_m})_G\big)$, which will be denoted by
${\mathcal E}$ again. 

Moreover, given two bitorsors ${\mathcal E}$ and ${\mathcal E}'$,
one defines their product by ${\mathcal E}\overset{{\mathbb
G}_m}\times {\mathcal E}'$, which is the quotient of ${\mathcal
E}\times {\mathcal E}'$ by the action of ${\mathbb G}_m$:
$$\begin{aligned} {\mathbb G}_m\times\big({\mathcal E}\times
{\mathcal E}'\big)&
\rightarrow {\mathcal E}\times {\mathcal E}' \\ (g,(e,e'))&\mapsto
(e\cdot g,g\cdot e')
\end{aligned}$$ (where the dot denotes the actions on ${\mathcal E}$
and ${\mathcal E}'$).

From \cite{SGA} \S1.3.4 we know that the group law of ${\mathcal
E}$ induces a canonical isomorphism:
\beq p_1^*{\mathcal E}\overset{{\mathbb G}_m}\times p_2^* {\mathcal
E}\,\iso\,m^* {\mathcal E}
\label{eq:square1}\end{equation} of $\big(({{\mathbb G}_m})_{G\times
G},({{\mathbb G}_m})_{G\times G}\big)$-bitorsors (where $p_i:G\times
G\to G$ is the projection in the
$i$-th component and $m$ the group law of $G$).

Conversely, a bitorsor ${\mathcal E}$ satisfying \ref{eq:square1}
and  an associative type property (see \cite{SGA} for the precise
statement) determines an extension of $G$. 

Observe that one can associate a line bundle to such an extension.
Given:
$$0\to {\mathbb G}_m \to {\mathcal E} \to G \to 0$$
consider the line bundle:
$$\L\,:=\,{\mathcal E}\overset{{\mathbb G}_m}\times {\mathbb A}_k^1$$
where ${\mathcal E}$ is interpreted as a principal fiber
bundle of group ${\mathbb G}_m$ and ${\mathbb G}_m$ acts on
${\mathbb A}_k^1$ by the trivial character and on ${\mathcal E}$ via
the inclusion ${\mathbb G}_m\subset {\mathcal E}$. Further, the
structure of ${\mathcal E}$ implies that there exists a canonical
isomorphism:
\beq
p_1^*\L\otimes p_2^*\L\,\iso\, m^*\L
\label{eq:square2}\end{equation}

One proves that the product of bitorsors corresponds to the tensor
product of line bundles; that is, for two extensions ${\mathcal E}$
and ${\mathcal E}'$ there exists a canonical isomorphism:
$$L_{{\mathcal E}\overset{{\mathbb G}_m}\times {\mathcal
E}'}\,\iso\, L_{\mathcal E}\otimes L_{{\mathcal E}'}$$

Conversely, if $\L$ is a  line bundle satisfying \ref{eq:square2} and
an associative type property, then the principal fibre bundle 
$\underline{\operatorname{Isom}}(\o_G,\L)$ is a principal fibre
bundle of group ${\mathbb G}_m$ which can be endowed with
the structure of central extension such that the associated line
bundle is $\L$.

\begin{defn}
The invertible sheaf on $G$ associated with $\widetilde G_\beta$ will
be denoted by $\Lambda_\beta$. 
\end{defn}

\section{Main Results}

\subsection{Modular properties of the $\tau$-function}\label{subsect:modular}

Let us fix a point $X\in\grv$ and a non negative integer $\beta$. 
From Theorem \ref{thm:glact} we know that there exists $L_\beta$, a
line bundle over $G$, such that:
\beq{\mu_\beta}_\bullet^*p_2^*\det_V \simeq
 p_2^*\det_V\otimes p_1^*L_\beta
\label{eq:inverseimage}\end{equation}
where:
 $$G\times \grv\overset{\mu_\beta}
\longrightarrow  G\times \grv\overset{p_2}\to\grv$$
Then, restricting to $G\times {X}$ and looking at sections we have:
$$\Omega_+(\mu_\beta(g)(X))\,=\,l_\beta(g)\cdot\Omega_+(X)$$
 for a certain section $l_\beta(g)$ of $L_\beta$ (we assume here that $\Omega_+(X)\neq 0$, so that it generates
$(\detd_V)_{X}$).

The above identity is the cornerstone of the modular properties of the $\tau$-functions.
However, let us give a more precise statement. Assume that the orbit of $X$
under
$\Gamma$ (consisting of invertible Laurent series acting by
multiplication, see \cite{AMP}) is  contained in $F_{V^+}$. Note, further,
that $L_\beta$ may be trivialized.  Then, with the above premises, the
following Theorem holds:

\begin{thm}
There exists a function $\bar l_\beta(g)$ on $G$, such that:
$$\tau_{\mu_\beta(g)(X)}\,=\,\bar l_\beta(g)\cdot \tau_X$$
\end{thm}

To finish this section let us offer a few hints on the explicit
computation of $l_\beta$. The previous statement is to be understood as an equality of $S$-valued functions (for a fixed $k$-scheme $S$ and $g\in G(S)$).

However, in order to describe this isomorphism explicitly it suffices to
deal with the case of the universal automorphism, ${\bf g}$, corresponding
to the identity point of $G(G)$. Note that the following relation holds:
$$\mu_\beta({\bf g})\,=\,{\bf g}'\circ \mu_{\beta-1}({\bf g})$$
(where ${\bf g}'$ acts as a homothety) and observe that the proof of
Theorem \ref{thm:glact} implies that the existence of canonical
isomorphisms:
$$\begin{aligned}
\mu_\beta({\bf g})_\bullet^*p_2^*\det_V& \simeq
\mu_{\beta-1}({\bf g})_\bullet^*p_2^*\det_V\otimes p_1^*(N) \qquad
\beta\geq1 
\\
\mu_0({\bf g})_\bullet^*p_2^*\det_V& \simeq
p_2^*\det_V\otimes p_1^*(M)
\end{aligned}$$
where:  
\begin{itemize}
\item $M=(\wedge \hat V^+_G/{\bf g}(\hat A_G))\otimes
 (\wedge \hat V^+_G\hat A_G)^*$,
\item $N=(\wedge \hat V^+_G/{\bf g}'\cdot \hat A_G)\otimes
 (\wedge \hat V^+_G/\hat A_G)^*$,
\end{itemize}
($A\in \B$ is locally choosen such that $A\subset V^+$, ${\bf g}'\cdot \hat
A_G\subset \hat V^+_G$ and
${\bf g}(\hat A_G)\subset \hat V^+_G$). Thus, we obtain:
\beq
L_\beta^*\,=\, M\otimes N^\beta
\label{eq:emeene}\end{equation}
and the computation of $\bar l_\beta(g):=l_\beta(g)/l_\beta(1)$ ($g\in
G(S)$) is now straightforward.

\begin{rem}
The above Theorem can be interpreted as the formal version of
Theorems~5.10 and~5.11 of \cite{KNTY}.
\end{rem}

\subsection{Central Extensions of $G$ and $\Lie(G)$}

Along the rest of this section it will be assumed that $k={\mathbb C}$.
Nevertheless, some results remain valid for $\char(k)=0$.
(We refer the reader to Appendix B for notations and the main results
on Lie theory for formal group schemes).

\begin{thm}\label{thm:exten-g-lieg}
The functor $\Lie$ induces an injective group homomorphism:
$$\ext^1(G,{\mathbb G}_m)\hookrightarrow \ext^1(\Lie(G),\widehat
{\mathbb G}_a)$$
\end{thm}

\begin{pf}
Here $\ext^1(G,{\mathbb G}_m)$ denotes the group of equivalence
classes of central extensions of $G$ by ${\mathbb G}_m$ as
formal groups, and $\ext^1(\Lie(G),\widehat {\mathbb G}_a)$
denotes the group of equivalence classes of central extensions of
Lie algebras.

Given an extension of $G$, $\widetilde G$, the restriction of the
group functors ${\mathbb G}_m, {\widetilde G}$ and
$ G$ to the category $\c_a$ ($\widehat {\mathbb G}_m, \widehat
{\widetilde G}$ and $\widehat G$ respectively) gives rise to a
class in $\ext^1(\widehat G,\widehat {\mathbb G}_m)$. Observe that
this map is injective. Recalling that
$\Lie(G)=\Lie(\widehat G)$, $\widehat {\mathbb G}_a\simeq \widehat
{\mathbb G}_m$ and Theorem \ref{ap:thm}, one concludes.
\end{pf}

\subsection{Some Canonical Isomorphisms}

\begin{thm}
$$L_\beta\,\simeq\, \Lambda_\beta$$
\end{thm}

\begin{pf}
Observe that equation \ref{eq:inverseimage} implies that:
$$p_1^*L_\beta\,\simeq\, 
Isom\big({\mu_\beta}_\bullet^*p_2^*\det_V,p_2^*\det_V\big)$$
and hence $L_\beta$ is the line bundle associated with the central
extension
$\tilde G_\beta$.
\end{pf}

\begin{thm}\label{thm:square} 
There are canonical isomorphisms:
$$m^*\Lambda_\beta\,\iso\, p_1^*\Lambda_\beta\otimes 
p_2^*\Lambda_\beta\qquad \forall \beta\in\Z$$
\end{thm}

\begin{pf}
This is a consequence of the subsection \ref{subsec:torsors}.
\end{pf}

\begin{thm}[{Local Mumford formula}]\label{thm:locmum}
There exist canonical isomorphisms of invertible sheaves:
$$\Lambda_\beta\,\iso\,\Lambda_1^{\otimes
(1-6\beta+6\beta^2)}\qquad\forall \beta\in\Z$$
\end{thm}

\begin{pf}
This is a consequence of Theorem \ref{thm:exten-g-lieg} and formula
\ref{eq:lie-mumford}.
\end{pf}

\begin{rem}
This Theorem is a local version of Mumford's formula. The next
subsection will throw some light on the relation between this formula and
the original global one. It is worth pointing out that the calculations
performed in subsection
\ref{subsect:modular} throw light on the explicit expression of the
above isomorphism. This can be done with  procedures similar to those of
\cite{BM}.
\end{rem}

\begin{cor}
Let $H$ be the subgroup of $G$ consisting of series $\sum_{i\geq
0}a_iz^i$ where $a_0$ is nilpotent and $a_1=1$.  

There is a canonical isomorphism:
$$(L_2\vert_H)^{\otimes 12}\,\simeq\, {\o_H}$$
(see \cite{Segal}~\S6 for explicit formulae). 
\end{cor}

\subsection{Orbits of $G$: relation with the moduli space of
curves}\label{subsec:modcur}

Recall from \cite{MP} the definition (which follows the ideas of
\cite{KNTY,U}):

\begin{defn}\label{5:defn:modinf} 
Set a $k$-scheme $S$. Define the
functor $\widetilde\M_g^{\infty}$ over the category of $k$-schemes by:
$$S\rightsquigarrow \widetilde\M_g^{\infty}(S)=
\{\text{ families $(C,D,z)$ over $S$ }\}$$  where these families
satisfy:
\begin{enumerate}
\item $\pi:C\to S$ is a proper flat morphism, whose geometric fibres
are integral curves of arithmetic genus $g$,
\item $\sigma:S\to C$ is a section of $\pi$, such that when
considered as a Cartier Divisor $D$ over $C$ it is smooth, of
relative degree 1, and flat over
$S$. (We understand that $D\subset C$ is smooth over $S$, iff for
every closed point $x\in D$ there exists an open neighborhood $U$ of
$x$ in $C$ such that the morphism $U\to S$ is smooth).
\item $\phi$ is an isomorphism of $\o_S$-algebras:
$$\widehat \Sigma_{C,D}\,\iso\, \o_S((z))$$
\end{enumerate}
\end{defn}

On the set $\widetilde\M_g^{\infty}(S)$ one can define an equivalence 
relation, $\sim$: $(C,D,z)$ and $(C',D',z')$ are said to be
equivalent, if there exists an isomorphism $C\to C'$ (over
$S$) such that the first family goes to the second under the induced
morphisms. Let us define the moduli functor of pointed curves of
genus $g$, $\M_g^{\infty}$, as the sheafication of
${\widetilde\M_g^{\infty}(S)}/{\sim}$. We know from Theorem~6.5  of
\cite{MP} that it is representable by a $k$-scheme $\M_g^{\infty}$. The
following Theorems are  now standard results:

\begin{thm}\label{thm:MP}
Let $g,\beta$ be two non-negative integer numbers. The ``Krichever
morphism'':
$$\begin{aligned}
K_{\beta}:\M_g^{\infty} &\longrightarrow\, \gr(k((t))(dt)^{\otimes
\beta})
\\ (C,p,z) &\longmapsto \, H^0(C-p,\omega_C^{\otimes \beta})
\end{aligned}
$$ 
is injective in a (formal) neighborhood of every geometric point. The
image will be denoted by $\M_{g,\beta}^{\infty}$.
\end{thm}

\begin{thm}\label{5:thm:mu0} 
The action $\mu_\beta$ of $G$ on $\grv$ induces an action
$\M_{g,\beta}^{\infty}$.
\end{thm}

\begin{pf} 
Recall that $G(R)=\aut_{R-alg}R((t))$ and that the points of
$\M_g^{\infty}(R)$ are certain sub-$R$-algebras of $R((t))$ ($R$ being a
commutative ring with identity). We thus have that the Krichever morphism
is equivariant with respect to the canonical action of $G$ on
$\M_g^{\infty}$ and $\mu_0$ on $\grv$. This implies the
$\beta=0$ case. The claim is now a straightforward generalization.
\end{pf}

In order to study the deformations of a given datum, more definitions
are needed. First, let $\M_g'$ be the subscheme of $\M_g^\infty$ defined
by the same conditions as in  Definition {\ref{5:defn:modinf}} except that
the third one is replaced by:
\begin{itemize}
\item $z$ is a formal trivialization of $C$ along $D$; that is, a
family of epimorphisms  of rings:
$$\o_C\longrightarrow
\sigma_*\left({\o_S[t]}/{t^m\,\o_S[t]}\right)\qquad m\in{\mathbb
N}$$ 
 compatible with respect to the canonical projections
${\o_S[t]}/{t^m\,\o_S[t]}\to {\o_S[t]}/{t^{m'}\,\o_S[t]}$ (for
$m\ge m'$), and such that that corresponding to $m=1$ equals
$\sigma$. 
\end{itemize}
Analogously, we introduce the
moduli space of pointed curves with an $n$-order trivialization,
$\M_g^n$ ($n\geq1$), as the $k$-scheme representing the
sheafication of the following functor over the category of
$k$-schemes:
$$S\rightsquigarrow \{\text{ families $(C,D,z)$ over $S$ }\}/\sim$$ 
where these families satisfy the same conditions except for the third
which is replaced by:
\begin{itemize}
\item $z$ is a $n$-order trivialization of $C$ along $D$; that is, 
an isomorphism:
$$\o_C/\o_C(-nD)\longrightarrow \sigma_*\left(\o_S[t]/t^n
\o_S[t]\right)$$
\end{itemize} 

The canonical projections $\M_g^{\infty}\to \M_g^n$
will be denoted by $p_n$. Observe that the natural projections
$\M_g^m\to \M_g^n$ ($m>n$) render $\{\M_g^n\vert n\geq 0\}$ an
inverse system and that $\M_g'$ is its inverse limit. In particular, we
have:
$$\M_g'\,=\,\limpl{n}\M_g^n$$

The deformation functor of a rational point $X$ of $\M_g^{\infty}$,
$D_X$, is the following functor over ${\mathcal C}_a$ (local
rational and artinian $k$-algebras):
$$A\,\rightsquigarrow \M_g^{\infty}(A)\underset
{\M_g^{\infty}(k)}\times\{X\}$$
Similarly, define $D_X'$ (resp. $D_{X_n}^n$), the deformation functor of
$X$ (resp. $X_n:=p_n(X)$) in $\M_g'$ (resp. $\M_g^n$). Since all the
$\M$'s are schemes, the corresponding deformation functors are
representable by the completion of the local rings.

\begin{lem}\label{5:lem5} 
Let $X\in \M_g'(k)$ be a triplet
$(C,p,z)$ with $C$ smooth. Then, the following sequence:
$$0\to H^0(C-p,{\mathbb T}_C)\,\to\,  
k((t))\partial_t \, \to\,  \limpl{n}H^1(C,{\mathbb
T}_C(-np))\to 0$$ 
(where ${\mathbb T}_C$ is the tangent sheaf on $C$)
is exact.
\end{lem}

\begin{pf} 
Let $m,n$ be two positive integers. Let us consider the
exact sequence:
$$0\to \o_C(-np)\to \o_C(mp)\to \o_C(mp)/\o_C(-np)\to 0$$ Since $z$
is a formal trivialization and $p$ is smooth, it induces an
isomorphism $\o_C(mp)/\o_C(-np)\iso t^{-m}k[[t]]/t^{n}k[[t]]$. 
Twisting the sequence with ${\mathbb T}_C$ and taking cohomology
one obtains:
$$\begin{gathered} 0\to H^0({\mathbb T}_C(-np))\to H^0({\mathbb
T}_C(mp))\to  t^{-m} k[[t]]\partial_t/t^n k[[t]]\partial_t \to
 \\  \to H^1({\mathbb T}_C(-np))\to H^1({\mathbb T}_C(mp))\to 0
\end{gathered}$$ since $\o_p\otimes_{\o_C}{\mathbb T}_C\simeq
<\partial_t>$. Taking direct limit on $m$ and inverse limit on $n$, the
result follows.
\end{pf}

\begin{thm}\label{5:thmmain} 
Let $k$ be a field of characteristic
$0$. Fix a rational point $X\in \M_g^{\infty}(k)$ corresponding to a
smooth curve. 

The morphism of functors:
$$\widehat G \,\longrightarrow\, D_X$$ induced by Theorem {\ref{5:thm:mu0}}
is surjective.  
\end{thm}

\begin{pf}
Let $\o_X$ be the local ring of $\M_g^\infty$ at $X$. 
The statement is equivalent to showing the surjectivity of the
induced maps:
$$\widehat G (A)\to D_X(A)=\sf(\widehat{\o}_X)(A)$$ for all
$A\in{\mathcal C}_a$. 
Now, Lemma {\ref{defor:lem}} reduces the problem to the case
$A=k[\epsilon]/\epsilon^2$:
$$\pi:\widehat G (k[\epsilon]/\epsilon^2)\to T_X{\M_g^{\infty}}$$
(where $T$ denotes the tangent space).

Observe that given $X$ there exists an element $g\in \widehat G$ such
that the transform of $X$ under $g$, $X^g$, belongs to $\M_g'$.  Then, the
proof is equivalent to showing that: 
$$T_{X^g}\M_g'\subseteq \im\pi$$

From Lemma {\ref{5:lem5}}, it follows that  the action of $\widehat G
(k[\epsilon]/\epsilon^2)= k((t))\partial_t =
\Lie(G)$ on $k((t))$ and that of $\der(H^0(C-p,\o_C))=H^0(C-p,{\mathbb
T}_C)$ on $H^0(C-p,\o_C)$ are compatible; further, the isotropy of $X$
under $k((t))\partial_t$ is precisely $H^0(C-p,{\mathbb T}_C)$. One can now
check that the above sequence induces a map:
$$ \limpl{n}H^1(C,{\mathbb T}_C(-np))\hookrightarrow T_X{\M_g^{\infty}}$$
whose image is naturally identified with $T_{X^g}\M_g'=\limpl{n}
T_{X_n}{\M_g^{n}}$ via the Kodaira-Spencer isomorphism. And the Theorem
follows.
\end{pf}

\begin{rem}
Let us now compare Theorem \ref{thm:locmum} and the standard
Mumford formula. Let $\M_g$ denote the moduli space of genus
$g$ curves, $\pi_g:{\mathcal C}_g\to \M_g$ the universal curve, and
$\omega$ the relative dualizing sheaf. Let us consider the family of
invertible sheaves:
$$\lambda_\beta\,:=\,\det(R^\bullet\pi_{g,\ast} \omega^{\otimes
\beta})\qquad \beta\in\Z$$

 Let $p:\M_g^\infty\to\M_g$ be the canonical projection. Then, it
holds that:
$$K_\beta^*\det_V\,\iso\, p^*\lambda_\beta$$

Furthermore, choose a rational point $X\in\M_g^{\infty}$ and let
$p_\beta$ be the composite:
$$\widehat G\,\to\, D_X^\beta \,\to\, \M_g $$
Then, it holds that there exist isomorphisms:
$$\Lambda_\beta\,\iso \, p_\beta^* \lambda_\beta$$
The compatibility of these isomorphisms with those of the Mumford
formula, $\lambda_\beta\iso\lambda_1^{\otimes(1-6\beta+6\beta^2)}$,
should follow from the proof Theorem \ref{thm:locmum} and the computations of 
\cite{BM,BS}.
\end{rem}


\section{Application to a non-perturbative approach to bosonic
strings}

Two standard approaches to Conformal Field Theories are based
on moduli spaces of Riemann Surfaces (with additional
structure) and on the representation theory of the Virasoro algebra,
respectively. It is thus natural to attempt to ``unify'' both
interpretations (e.g.
\cite{KNTY}). 

In our setting, subsections~\ref{subsect:lie} and~\ref{subsec:modcur}
unveil the important role of the group $G$ in both approaches. Motivated
by this fact and by the suggestions of \cite{BR} and \cite{Mo}, we propose
 $G$ as a ``universal moduli space'' which will allow 
formulation of a non-perturbative string theory. Let us remark that in the
formal geometric setting developed in \cite{MP}, the group $G$ is the
moduli space of formal curves.

Let us sketch how this construction should be carried out, although 
details and proofs will be given in a forthcoming paper.

Let us consider the vector space $V_d=\C^{d}\otimes_{\C}\C((t))$. The
natural representation, $\mu_1$, of $G$ on $V_1$ induces a
representation of $G$ on $V_d$, given by
$\mu_1\oplus\overset{d}\ldots\oplus\mu_1$. Following the procedure given
in \S~\ref{sect:G}, it is easily proved that this representation yields
an action, $\rho_d$, of $G$ on the Grassmannian $\gr(V_d)$ preserving the
determinant bundle. The corresponding central extension determines  a line
bundle $\L_{\rho_d}$ on $G$ with a bitorsor structure. 

In order to clarify the physical meaning of this higher dimensional
picture, it is worth pointing out that the Fock space corresponding to
string theory in the space-time
$\R^{2d-1,1}$ is naturally interpreted as a subspace of
$H^0(\gr(V_d),\det^*)$, the space of global sections of the
dual of the determinant bundle. Moreover, the actions of the Virasoro
algebra on the Fock space and that of $\Lie(G)$ on $H^0(\gr(V_d),\det^*)$
are compatible. 

The calculations in section~5 of \cite{BR} can now be restated in the
following form: there exists a canonical isomorphism of invertible sheaves:
$$\L_{\rho_d}\,\iso\, \Lambda_1^{\otimes d}$$
This isomorphism, together with the Local Mumford Formula (Theorem 
\ref{thm:locmum}), implies that $\L_{\rho_d}$ and $\Lambda_2$ are
isomorphic if and only if
$d=13$ (complex dimension).

Observe that the group scheme $\widetilde G$ carries a filtration
$\{G_n\vert n\geq 0\}$, where:
$$G_n(R)\,:=\, \{\phi\in\widetilde G(R)\vert \phi(t)=\sum_{i\geq
-m}a_it^{i} \text{ with }m\leq n\}$$
The restriction homomorphisms:
$$j_n^\ast : H^0(G,\Lambda_\beta)\to H^0(G_n,\Lambda_\beta\vert_{G_n})$$
associated with the inclusions $j_n:G_n\hookrightarrow G$ give:
$$j^\ast : H^0(G,\Lambda_\beta)\to
\limpl{n}H^0(G_n,\Lambda_\beta\vert_{G_n})$$

 Let $X$ be a rational point of $\M_g^\infty$. The action of $G$ on $X$
induces:
$$\phi_g^n:G_n\longrightarrow \M_g^\infty$$
which takes values in the deformation functor of $X$, $D_X$. Moreover,
$G_n\to D_X$ happens to be surjective for all $n\geq 3g-3$ (see
Theorem \ref{5:thmmain}). Denote by
$F_g\in H^0(G_{3g-3},\Lambda_2\vert_{G_{3g-3}})$ the inverse image by
$\phi_g^{3g-3}$ of the section of $\lambda_2$ corresponding to the
partition function of genus $g$. Then, there exists a global section $F\in
H^0(G,\Lambda_2)$ such that $j^*(F)$ is precisely $\{F_g\}$. 

The relationship between hermitian forms on the
canonical sheaf of a complex manifold and holomorphic measures on
them is well known. The generalization of this relation to
infinite-dimensional manifolds would allow us to give a genus-independent
Polyakov measure on $G$ constructed in terms of the above introduced $F$.

\appendix
\section{Deformation Theory}

Let us recall some notations and give some results on deformation
theory as exposed in \cite{Schl}.

Let $\c_a$ be the category of local rational Artin $k$-algebras. An
admissible linearly topologized $k$-algebra $\o$ (see
\cite{EGA}~\S7) canonically defines  a functor from $\c_a$ to the
category of sets:
$$A\,\rightsquigarrow h_{\o}(A):=\hom_{\text{cont}}(\o,A)$$
(where $A$ is endowed with the discrete topology). Observe that
$h_{\o}(A)=\hom_{\text{$k$-alg}}(\o,A)$ for a discrete $k$-algebra
$\o$. 

The condition that $h_{\o}$ consists of only one point is
equivalent to saying that $\o$ is local and rational. 

The definition below is that given in \cite{Schl}~2.2, which
generalizes the concept of ``formal smoothness'' of \cite{MatCA}.

\begin{defn}
A functor homomorphism $F\to G$ is smooth iff the morphism:
$$F(B)\,\to\,F(A)\times_{G(A)}G(B)$$
is surjective for every surjection $B\to A$ in $\c_a$.
\end{defn}

\begin{rem}
The following remarks merit attention:
\begin{itemize}
\item if $F\to G$ is smooth, then $F(A)\to G(A)$ is surjective for
all $A$ in $\c_a$ (\cite{Schl}~2.4),
\item $h_{\o}\to h_{\o'}$ is smooth iff $\o$ is a series power ring
over $\o'$ (\cite{Schl}~2.5),
\item $h_{\o}$ is said to be smooth  iff the canonical morphism
$h_{\o}\to h_{k}$ is smooth.
\end{itemize}
\end{rem}

The tangent space to  a functor
over $\c_a$, $F$, is defined by:
$${\frak t}_F\,:=\, F(k[\epsilon]/\epsilon^2)$$
Recall Lemma~2.10 of \cite{Schl}: if it holds that:
$$F(k[V\oplus W])\,\simeq \, F(k[V])\times F(k[W])$$
for arbitrary vector spaces $V,W$ (where $k[V]$ is the ring
$k\oplus V$ with $V^2=0$), then $ F(k[V])$ (and in particular
${\frak t}_F$) has a canonical vector space structure such that $
F(k[V])\simeq {\frak t}_F\otimes V$.
Observe that the functor $h_{\o}$ satisfies the above condition for
all $\o$.

\begin{lem}\label{defor:lem}
Let $\phi:F:=h_{\o_F}\to G:=h_{\o_G}$ and $F\to h_k$ be two morphisms
of functors over $\c_a$ such that:
\begin{itemize}
\item $F\to h_k$ is smooth,
\item the sets $F(k)$ and $G(k)$ consist of one element,
\item ${\frak t}_F:=F(k[\epsilon]/\epsilon^2)\to
{\frak t}_G:=G(k[\epsilon]/\epsilon^2)$ is surjective, 
\end{itemize}
then $F\to G$ is smooth (and hence surjective).
\end{lem}

\begin{pf}
First, we claim that $F(k[V])\to G(k[V])$ is surjective for
every $k$-vector space $V$ ($k[V]$ denotes the ring $k\oplus V$ in
which $V^2=0$). Since $F(k[V\oplus W])\simeq F(k[V])\times
F(k[W])$ and $G(k[V\oplus W])\simeq G(k[V])\times G(k[W])$ for
vector spaces $V,W$, Lemma~2.10 of
\cite{Schl} holds, and hence there are canonical vector space structures
on $F(k[V])$ and $G(k[V])$ such that they are isomorphic to ${\frak
t}_F\otimes V$ and ${\frak t}_G\otimes V$ (in a functorial way)
respectively. Since ${\frak t}_F\to {\frak t}_G$ is surjective by
hypothesis, the claim follows.

Let $A$ be an object of $\c_a$ and $I\subset A$ an ideal such that
$I^2=0$. Then, one has a commutative diagramm:
$$\CD
F(A) @>\phi_A>> G(A) \\
@V\rho_F VV @V\rho_G VV \\
F(A/I) @>\phi_I>> G(A/I)
\endCD$$
where we can assume by induction over $\dim_k A$ that $\phi_{I}$ is
surjective (since ${\frak t}_F\to {\frak t}_G$ is surjective). 

Let $(f,g)$ be an element of $ F(A/I)\times
G(A)$ such that $\phi_I(f)=\rho_G(g)$. Since $F\to
h_k$ is smooth and $\o$ is local it follows that $\rho_F$ is a
surjection. Let $\bar f\in F(A)$ be a preimage of
$f$.  Then the images of $\phi_A(\bar f)$ and $g$ under $\rho_G$
coincide; both of them are $\phi_I(f)$. Note that
$\rho_G^{-1}(\phi_I(f))$ is an affine space modeled over
$\der_k(\o_G,I)$; or what amounts to
the same:
$$g-\phi_n(\bar f)\in \der_k(\o_G,I)$$

Observe that the bottom arrow of the following commutative diagramm:
$$\CD 
\der_k(\o_F,I) @>>> \der_k(\o_G,I) \\
@V\simeq VV @V\simeq VV  \\
F(k[I]) @>>> G(k[I])
\endCD$$
is surjective. Let $D\in F(k[I])$ be a preimage of $g-\phi_n(\bar
f)$.

It is now easy to verify that $\bar f+D$ is a preimage of $(f,g)$
under the induced morphism:
$$ F(A)\longrightarrow F(A/I)\times G(A)$$
and the statement follows.
\end{pf}

\section{Lie Theory}\label{sect:Lie}

This appendix aims at generalizing some results of Lie Theory for
the case of (infinite) formal groups. To this end, we 
recall some more results of \cite{Schl} and proceed with ideas
quite close to those of \cite{Haz}~\S14.

\begin{defn}
A functor $F$ from $\c_a$ to the category of groups will be called 
a group functor. If, moreover, there exists a $k$-algebra $\o$ and an
isomorphism $F\simeq h_{\o}$, then $F$ will be called  a formal
group functor. 
\end{defn}

Let $\c_{\text{gr}}$ and  $\c_{\text{for gr}}$ denote the categories
of group functors and formal group functors over $\c_a$,
respectively. Let $\c^0_{\text{for gr}}$ denote the full subcategory
of $\c_{\text{for gr}}$ consisting of those $F$ such that $F(k)$ has
only one element and $F$ is smooth.

\begin{rem}\hfill
\begin{itemize}
\item Let $F$ be a formal group functor over $\c_a$. Then, the
``tangent space at the neutrum'':
$$\Lie(F)\,:=\, F(k[\epsilon]/\epsilon^2)\times_{F(k)}\{1\}$$
(which coincides with ${\frak t}_F$) is a Lie algebra where
the Lie bracket is induced by the product of $F$.
\item Finally, for a formal group functor and a morphism $A\to A/I$
with $I^2=0$ one has the following exact sequence of groups:
$$0\to F(k[I])\to F(A)\to F(A/I)\to 0$$
\end{itemize}
\end{rem}

\begin{lem}\label{ap:lem:1}
Let $\char(k)=0$. Let $F$ and $G$ be two formal group
functors. Assume that $F$ is smooth and that $F(k)=\{e\}$ (one
point). Then, the canonical map:
$$\hom_{\text{gr}}(F,G)\to \hom_{\text{vect. sp.}}({\frak
t}_F,{\frak t}_G)$$
is injective.
\end{lem}

\begin{pf}
Let $A$ be an object of $\c_a$ and ${\frak m}\subset A$ its maximal
ideal and $n$ such that ${\frak m}^{n+1}=0$. Let $\phi,\psi$ be in 
$\hom_{\text{gr}}(F,G)$ such that the induced
vector space homomorphisms $\phi_*,\psi_*$ from ${\frak
t}_F$ to ${\frak t}_G$ coincide. One has to prove that $\phi=\psi$.

Let us first deal with the case $n=1$. By Lemma~2.10 of
\cite{Schl}, there exist functorial isomorphisms $F(A)\simeq {\frak
t}_F\otimes {\frak m}$ and $G(A)\simeq {\frak
t}_G\otimes {\frak m}$ ($\frak m$ as a $k$-vector space). It is now
clear that both, $\phi$ and $\psi$, give the same morphism $F(A)\to
G(A)$.

Now assume $n\geq 2$. Using the Nakayama Lemma one obtains a surjection:
$$A_{r,n}:=k[x_1,\ldots,x_r]/(x_1^{n+1},\ldots,x_r^{n+1})\to A$$
and hence a commutative diagramm:
$$\CD F(A_{r,n}) @>>> F(A) \\
@V\phi VV @V\phi VV \\
G(A_{r,n}) @>>> G(A)
\endCD$$
and similarly for $\psi$. Observe that the top row is surjective
since $F$ is smooth and $A_{r,n}\to A$ is surjective.
Therefore, it suffices to prove the statement for $A_{r,n}$.

Note that the injection  $A_{r,n}\hookrightarrow A_{r\cdot n,1}$
($\char(k)=0$):
$$\begin{aligned}
k[\{x_{i}\,\vert \, 1\leq i\leq r\}]/(x_i^{n+1})
&\to k[\{x_{ij}\,\vert \, 1\leq i\leq r, 1\leq j\leq n\}]/(x_{ij}^2)
\\
x_i\quad&\longmapsto x_{i1}+\ldots+ x_{in}
\end{aligned}$$
induces two commutative diagramms (for $\phi$ and $\psi$):
$$\CD
0 @>>> F(A_{r,n}) @>>> F(A_{r\cdot n,1}) \\
@. @VVV @VVV \\
0 @>>> G(A_{r,n}) @>>> G(A_{r\cdot n,1}) 
\endCD$$
It is then enough to check the case of $A_{r,1}$. Let us
proceed by induction on $r$. The case $r=1$ follows directly from
the hypotheses.

We claim that the the following 
diagramm is commutative:
$$\CD 
0@>>> F(k[\ker(p)]) @>>> F(A_{r,1}) @>p_F>> F(A_{r-1,1}) @>>> 0 \\
@. @V\phi_p VV @V\phi_r VV @V\phi_{r-1} VV \\
0@>>> G(k[\ker(p)]) @>>> G(A_{r,1}) @>p_G>> G(A_{r-1,1}) @>>> 0
\endCD$$
(and analogously for $\psi$). The morphisms $p_F$ and $p_G$ are
surjective since they have sections, because the natural inclusion
$A_{r-1,1}\hookrightarrow A_{r,1}$ is a section of the projection:
$$\begin{aligned}
p:A_{r,1} &\to A_{r-1,1} \\
x_r &\mapsto 0
\end{aligned}$$
Bearing in mind that $(\ker p)^2=0$, the claim follows.

The first case, which we have already proved (the square of the maximal
ideal is $(0)$), implies that the $\phi_p =\psi_p$. The induction's
hypothesis implies that $\phi_{r-1}=\psi_{r-1}$.

Now, recalling that both sequences split, one concludes that
$\phi_r=\psi_r$ as desired.
\end{pf}

Let us now relate the study of group functors with that of Lie
algebras. Let $\c_{Lie}$ denotes the category of Lie
$k$-algebras. Then, there is a functor:
$$\begin{aligned}
\Lie:\c_{\text{for gr}} &\longrightarrow \c_{\text{Lie}} \\
F &\longmapsto  \Lie(F)={\frak t}_F
\end{aligned}$$

For a Lie $k$-algebra ${\frak L}$ define a functor on $\c_a$:
$$A\,\rightsquigarrow \underline{\frak L}(A):={\frak L}\otimes_k
{\frak m}_A$$ (the Lie bracket of $\underline{\frak L}(A)$ is that
of ${\frak L}$ extended by $A$-linearity). 

Let $CH(x,y)$ denote the Campbell-Hausdorff series
(see, for instance, \cite{Haz}~14.4.15):
{\small\beq
CH(x,y)\,=\,
x+y+\frac12[x,y]+\frac1{12}[x,[x,y]]+\frac1{12}[y,[y,x]]+\ldots
\label{eq:CH}\end{equation}}
then the map:
$$\begin{aligned}
\underline{\frak L}(A)\times \underline{\frak
L}(A) & \to \underline{\frak L}(A) \\
(x,y)\,&\longmapsto CH(x,y)
\end{aligned}$$
(note that $CH(x,y)$ is a finite sum since $A$ is artinian) endows
$\underline{\frak L}(A)$ with a group structure
(\cite{Haz}~14.4.13-16). Let us denote this group functor  by
${\frak L}^g$. Moreover
${\frak L}^g\to h_k$ is smooth and ${\frak L}^g(k)$ consists of one point.
Finally, since
$CH(x,y)$ only depends on additions of iterated Lie brackets one has
that every morphism of Lie algebras ${\frak L}_1\to {\frak L}_2$
induces a morphism of group functors ${\frak L}_1^g\to
{\frak L}_2^g$. In other words, there is a functor:
$$\begin{aligned}
{\frak G}:\c_{\text{Lie}} &\longrightarrow \c_{\text{gr}} \\
{\frak L} &\longmapsto  {\frak L}^g
\end{aligned}$$
such that $\Lie\circ {\frak G}=\id$.

\begin{exam}
It is now easy to prove that finite dimensional Lie algebras are the
Lie algebras of formal groups. Indeed, let ${\frak L}^*$ be the
dual vector space of a given Lie algebra ${\frak L}$. Then, it holds
that:
$$\hom_{\text{cont}}(\o,A)\,=\,
{\frak L}\otimes_k {\frak m}_A$$
where $\o:=\widehat S^\bullet {\frak L}^*$ is the completion of the
symmetric algebra, $S^\bullet {\frak L}^*$, with respect to the
maximal ideal generated by ${\frak L}^*$.

It is now straightforward to see that ${\frak L}^g=h_{\o}$ and that:
$$\Lie(h_\o)=({\frak m}_\o/{\frak m}^2_\o)^*={\frak L}$$
\end{exam}

\begin{lem}
Let $F$ be an object of $\c^0_{\text{for gr}}$. The
functor homomorphism (which will be called exponential) defined by:
$$ \begin{aligned}
\underline{\frak t}_F &\to F \\
D &\mapsto exp(D):=\sum_{i\geq 0}\frac{1}{i!}D^i
\end{aligned}$$
yields an isomorphism ${\frak t}^g_F\simeq F$.
\end{lem}

\begin{pf}
Note that the sum is finite since $D\in \underline{\frak
t}_F(A)={\frak t}_F\otimes {\frak m}_A$ (for $A\in\c_a$) is of the
type $D=\sum_j m_j D_j$ (where $m_j\in{\frak m}_A$ and $D_j\in
{\frak t}_F$) and hence $D^i$ has coefficients in ${\frak m}_A^j$.
By the above construction, the exponential is a group homomorphism
since it holds that (\cite{Haz}~14.14):
$$exp(D)\cdot exp(D')\,=\, exp(CH(D,D'))$$

In the same way that the exponential map has been defined  a logarithm can
also be introduced. Now the conclusion follows trivially.
\end{pf}

From all these results one has the main Theorem of this appendix
which is a version for (certain) non-commutative group functors of the
standard Lie Third Theorem.

\begin{thm}\label{ap:thm}
The functor $\Lie$ renders $\c^0_{\text{for gr}}$ a full
subcategory of $\c_{Lie}$.
\end{thm}

\begin{pf}
This follows from the following two facts:
\begin{itemize}
\item if $F,G\in\c_{gr}$ have isomorphic Lie algebras ${\frak
t}_F\simeq {\frak t}_G$, then they are isomorphic. (Recall that there
are group isomorphisms ${\frak t}_F^g\simeq F$ and ${\frak
t}_G^g\simeq G$).
\item $\hom_{\c_{gr}}(F,G)\simeq \hom_{\c_{Lie}}({\frak t}_F,{\frak
t}_G)$ (Lemma \ref{ap:lem:1} proves the injectivity and the
equality  $\Lie\circ {\frak G}=\id$ the surjectivity).
\end{itemize}
\end{pf}


\end{document}